\documentclass[final]{svjour3}
\usepackage{gensymb}
\usepackage{graphicx}
\usepackage{rotating}
\usepackage{amssymb}
\usepackage{mathptmx}
\usepackage[numbers]{natbib}
\makeatletter
\journalname{Journal of Low Temperature Physics}

\bibpunct{}{}{,}{s}{}{,}

\begin{document}

\newcommand{\hdblarrow}{H\makebox[0.9ex][l]{$\downdownarrows$}-}
\title{Simulating Cosmic Microwave Background anisotropy measurements for Microwave Kinetic Inductance Devices.
}

\author{R. Basu Thakur, J. Henning, P. S. Barry, E. Shirokoff, Q. Y. Tang}

\institute{Kavli Institute for Cosmological Physics, Department of Astronomy and Astrophysics at the University of Chicago, Chicago, IL, USA 60637
\email{ritoban@uchicago.edu}}

\maketitle

\begin{abstract}

 Microwave Kinetic Inductance Devices (MKIDs) are poised to allow for massively and natively multiplexed photon detectors arrays and are a natural choice for the next-generation CMB-Stage 4 experiment which will require $10^5$ detectors. In this proceeding we discuss what noise performance of present generation MKIDs implies for CMB measurements. We consider MKID noise spectra and simulate a telescope scan strategy which projects the detector noise onto the CMB sky. We then analyze the simulated CMB + MKID noise to understand particularly low frequency noise affects the various features of the CMB, and thusly set up a framework connecting MKID characteristics with scan strategies, to the type of CMB signals we may probe with such detectors.


\end{abstract}

\section{Introduction}

The Stage-4 Cosmic Microwave Background (CMB-S4) collaboration plans to utilize $\sim10^5$ low-temperature low-noise detectors, to measure the CMB with $\lesssim 1$nK sensitivity across angular scales from $10^{-1}$ to $10^{2}$ degrees, Ref.[1]. These measurements will provide extremely precise constraints on the matter-energy budget and dynamics of the universe, and probe several interesting early universe inflationary theories. Microwave Kinetic Inductance Devices (MKIDs)  are naturally suited to perform such ambitious surveys. MKIDs are superconducting resonators whose resonance frequencies shift in proportion to incident power. These resonators thus enable multiplexing large numbers of photon detectors, in principle improving the current multiplexing factor for CMB experiments by over an order of magnitude. In sub-millimeter astronomy, several groups have demonstrated the viability of MKIDs with respect to both multiplexing factors as well as noise requirements, these include SuperSpec, NIKA2 and BLAST-TNG, Refs.[2-6]\\

To obtain unbiased high-precision cosmological measurements we must consider not only the absolute noise-floor of the future detectors but also their spectral shape, since cosmological parameter derivation relies on the slope and relative heights of features of the CMB spectrum. We address this issue by constructing a framework that directly connects laboratory measurements of detector noise to the cosmological parameters that we ultimately seek. Our framework is applicable to any cryogenic detector technology and in this article we use MKIDs as a specific example.

\section{Noise, measurements and models}

\begin{figure}[h!!]
\begin{center}
\includegraphics[width=0.7\textwidth]{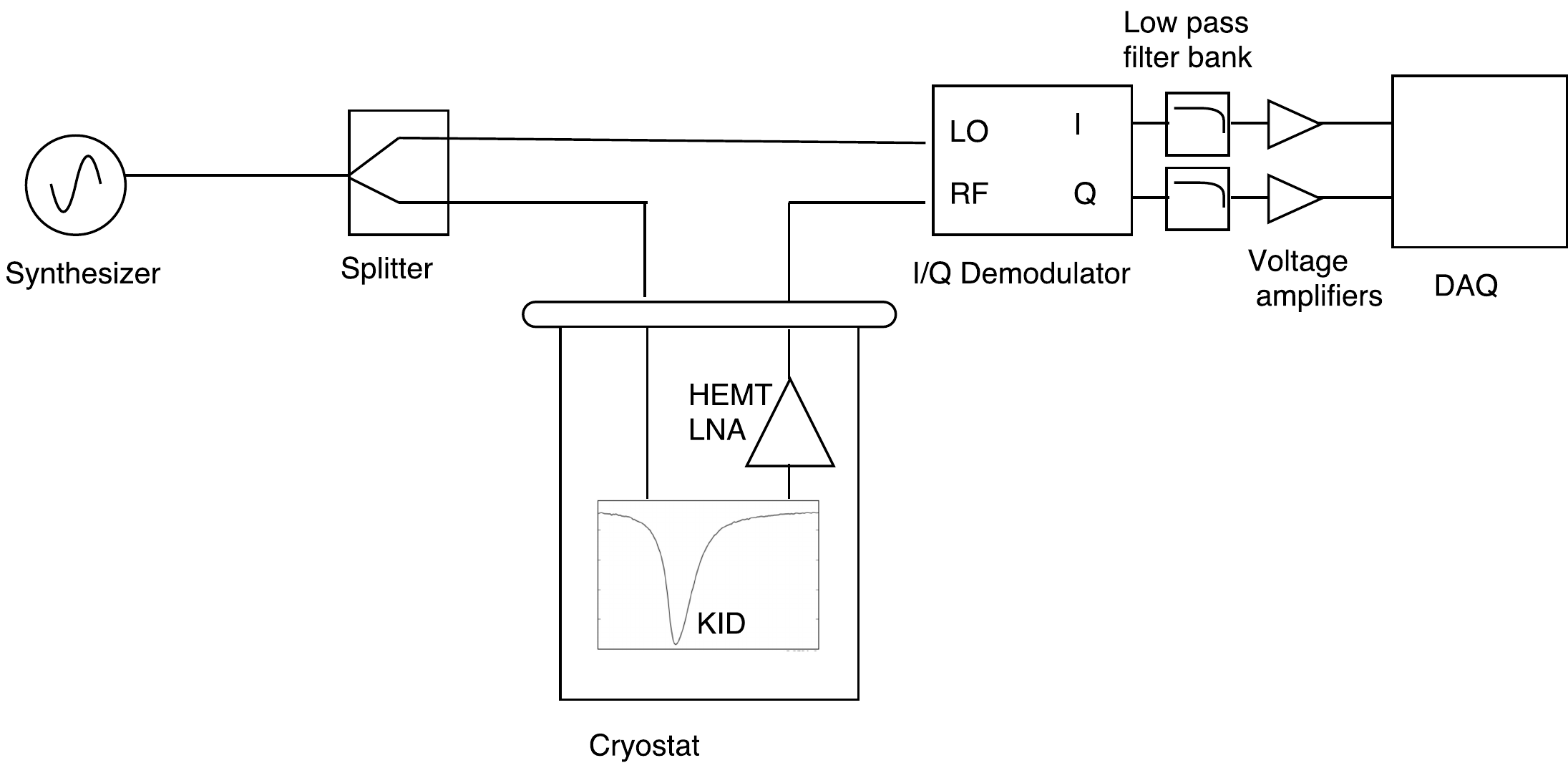}
\caption{Single tone KID noise measurement setup: a synthesizer produces a single microwave tone which goes to a splitter. One-half of the synthesizer power enters the cryostat while the other half goes to the local oscillator (LO) input of a quadrature demodulator. The microwave signal entering the cryostat drives the superconducting resonator (S21 profile of an MKID is shown in the schematic), whose output is amplified cryogenically via a low noise amplifier (HEMT or Si-Ge amplifiers) before it exits the cryostat. This RF output then enters the demodulator where it is demodulated by the LO resulting in In-phase (I) and Quadrature (Q) signals which are filtered, amplified and digitized.  }
\end{center}
\label{Fig:Ro}
\end{figure}

Noise measurement for KID arrays may be done with single-tone or multi-tone methods Ref [2-4]. Outline of single tone read out is as follows: a single microwave tone from a synthesizer is input to a KID, where we choose the frequency to be either on resonance or off resonance. The output is amplified, at 4K and further at 300K, and this output signal is demodulated with respect to the synthesizer input. In presence of physical processes such as two-level-system noise, which give rise to low-frequency fluctuations, and thermal generation-recombination of quasiparticles, we obtain both phase and amplitude noise, typically over a bandwidth of $<$ 100 kHz. This is schematically shown in Fig.~\ref{Fig:Ro}\\

\begin{figure}[h!!]
\begin{center}
\includegraphics[width=0.6\textwidth]{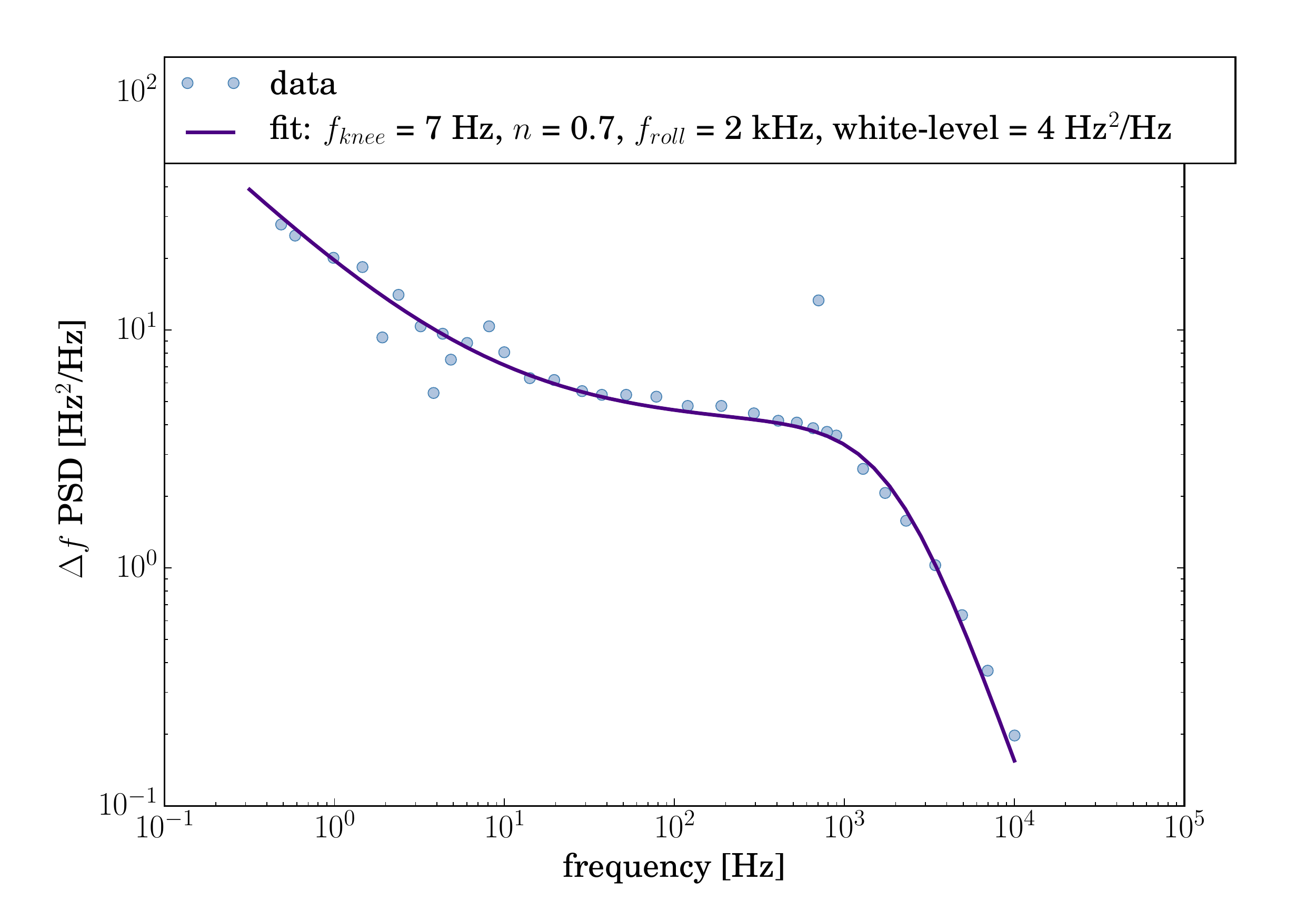}
\caption{ An example power spectral density (PSD) of a 350 GHz lumped element KID. The measurement is the frequency shift of a resonator with time, which we Fourier transform to obtain the PSD, hence units of Hz$^2$/Hz .The PSD is fit very well by the model described in Eqn.~\ref{eq:eqSf}, and the noteworthy features are the 1/f like low-frequency noise and the white noise level at higher frequencies rolled-off by design. }
\end{center}
\label{Fig:psdfit}
\end{figure}

A typical measurement of the noise power spectral density is shown in Fig.~\ref{Fig:psdfit}, and this is the noise on resonance, i.e., the nominal configuration for an optically coupled KID. The white-noise or the noise equivalent power for a photon-noise limited detector is characterized as ${NEP}^2 = 2h\nu_0 P_{opt}+\zeta P_{opt}^2/\Delta \nu$, where $\nu_0, \Delta \nu$ define the band-center and band-width of the spectrum seen by the detector, $P_{opt}$ is the total optical power incident in this band, and $\zeta$ is related to the optical efficiency and the number of photon modes around $\nu_0$. A slightly higher $NEP^2$, perhaps due to generation-recombination noise, can be proportionally compensated by increasing the sampling time and the number of detectors. Indeed for CMB S4, the 1nK sensitivity and $10^5$ detector count are related by assumptions of scan strategies and NEPs extrapolated from present experiments.\\

For typical KIDs the shape of the noise power spectral density (PSD) is characterized by the fit to the data shown in Fig.~\ref{Fig:psdfit}, namely there is a low-frequency noise (called 1/f noise) in addition to the white-noise level, and there is a roll-off as a result of readout design. While every KID will have its own PSD, we define a generic model $S(f)$ strictly as a function the low-frequency characteristics, namely the power of the 1/f term $n$, and the frequency $f_{knee}$ where the low-frequency power matches white noise level $w$. An artificial roll off is introduced with $f_{roll} \gg f_{knee}$ fixed at 1 kHz, see Eqn.~\ref{eq:eqSf}
\begin{equation}
	S(f \, | n,f_{knee}) = w \frac{1+\left(f/f_{knee}\right)^{-n}}{1+(f/f_{roll})^2}
\label{eq:eqSf}
\end{equation}

The final goal of device fabrication is to have the lowest possible values of $n,f_{knee}$, however at present most devices demonstrate $n \sim O(1)$ and $f_{knee} \sim O(1Hz)$. 

\section{Sky projection}
\subsection{Noise Maps}
As a telescope scans the sky, the CMB is realized as a sum of spatial modes of the temperature anisotropy. A spatial mode of wavelength $\theta_i$ (degrees-on-the-sky) is mapped as a temporal mode with frequency $f_i \approx v_s / \theta_i$ where $v_s$ is the scan speed in $^{\circ}$/s. Thus a 1$^{\circ}$ dipole, for a scan-speed $\sim 1^{\circ}$/s, results in a 1 Hz sine-wave in time stream. This numerology is noteworthy since the first acoustic peak of the CMB is seen as structures extending $\sim 1^{\circ}$ on the sky, and for many KIDs 1/f noise is seen to be present at $O(1)$ Hz, see Fig.~\ref{Fig:psdfit}. \\

In our studies we consider realizations of the CMB in spatial coordinates and project on to this a noise map. The noise maps are derived from constructing noise time streams and then modulating the time streams into spatial modes using a scan speed $v_s = 1^{\circ}/s$. For a given PSD, $S(f \, | n,f_{knee})$, we may generate $N_n$ random noise samples following Eqn.~\ref{eq:nkt}.

\begin{equation}
n^{(k)}(t_j) = \sum_i \sqrt{S(f_i | n,f_{knee})} \cos{(2\pi f_i t_j + \phi^{(k)}(f_i))} \frac{\Delta f}{\sqrt{f_s}/2}, \; k \in (1 \ldots N_n)
\label{eq:nkt}
\end{equation}

Eqn.~\ref{eq:nkt} is effectively the PSD's discrete inverse Fourier transform, i.e., $\langle \tilde{n}^{*(k)}\tilde{n}^{(k)} \rangle_{(k)} = IFFT\{PSD\}  $. A PSD does not have phase information, thus we inject random phases for every noise realization via $\phi^{(k)}(f_i)$. For generic phase-coherent noise we explicitly allow frequency dependent phase. However, hereon we make the standard simplification of phase incoherency, for every frequency $f_i$ we choose $\phi^{(k)}$ randomly from the uniform distribution $\mathcal{U}(0,2\pi)$. Finally in Eqn.~\ref{eq:nkt}, $\Delta f, f_s$ stand for the bin-width of the discrete frequency and sampling frequency respectively.

\begin{figure}[h!!]
\begin{center}
\includegraphics[width=0.62\textwidth]{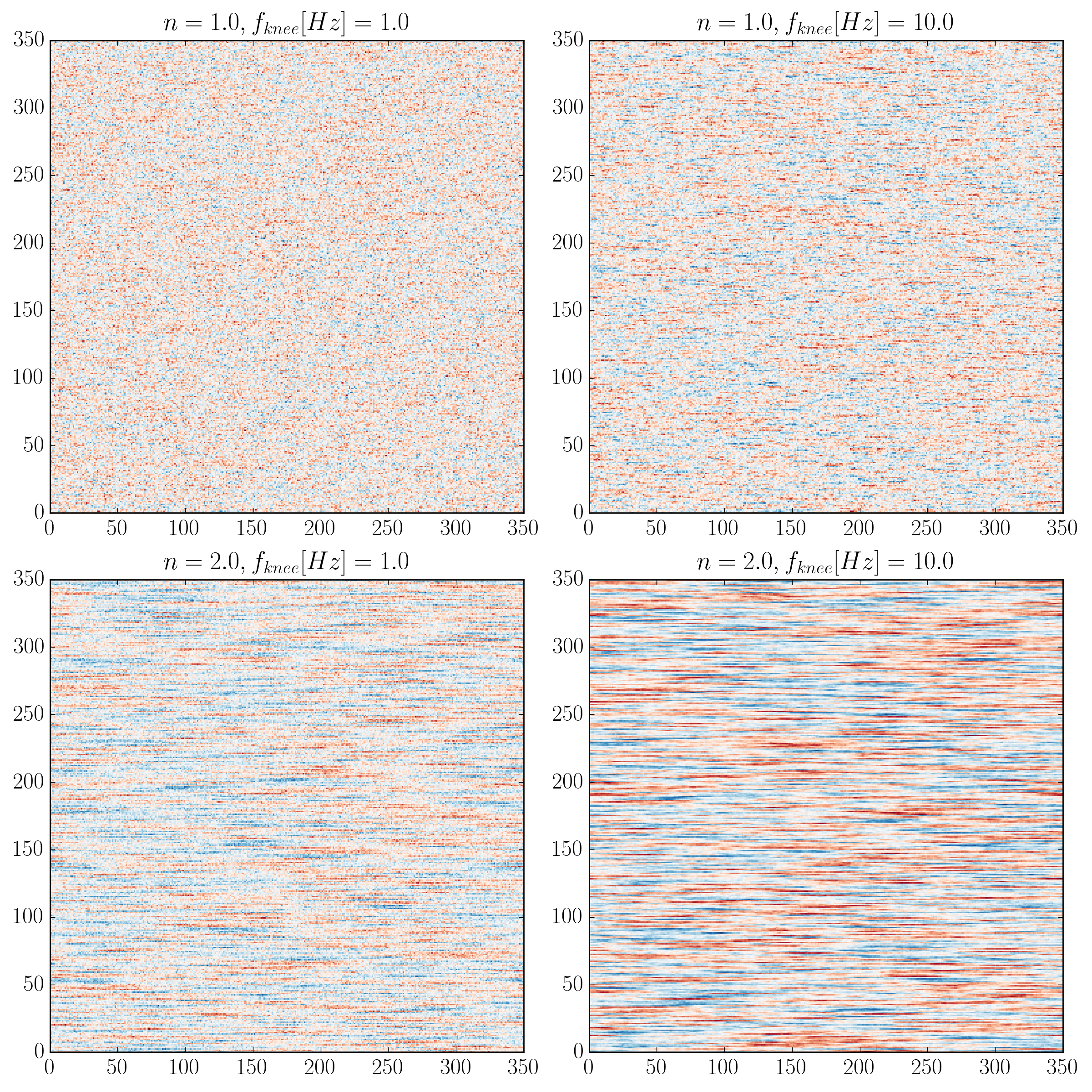}
\caption{Examples of four noise maps: plots are titled with the values of $n,\; f_{knee}$ used in each case. We clearly note how excess 1/f noise leads to {\it fake} hot-and-cold (red/blue contrast) spots which will mask the true stochastic hot-cold fluctuations of the CMB.  The coldest spots have a noise depth $\sim 10 \mu K$arcmin. In these maps, the raster scan happens in the horizontal or azimuthal direction.}
\end{center}
\label{Fig:noismaps}
\end{figure}

We projected these time streams into spatial modes and construct noise maps as shown in Fig.~\ref{Fig:noismaps}, which are added to signal (CMB temperature) maps. The CMB maps are  $N_g \times N_g$ pixels each pixel being $\theta_{res}$ (angular resolution) in size. We model raster scans: the telescope scans in azimuth with scan speed $v_s$ and then steps up slightly in elevation, scanning again in azimuth, repeating this pattern until it has scanned $(N_g\theta_{res})^2$ square-degrees. For the noise realizations we choose sampling frequency $f_s > v_s/\theta_{res}$ and the total time for one noise time-stream is $T > N_g\theta_{res}/v_s$. Each time stream is binned with bin-width $\Delta t = \theta_{res}/v_s$, and the noise which has $\Delta t f_s$ samples per bin, is averaged to assign a noise-temperature value in that pixel. These pixelated noise time streams are stacked following the raster scan pattern to generate the noise maps, see Fig.~\ref{Fig:noismaps}. We immediately note how different noise PSDs may introduce temperature anisotropies, and how such noisy features might qualitatively affect the measurement of the CMB. 

\subsection{Signal+Noise spectra}

\begin{figure}[h!!]
\begin{center}
\includegraphics[width=0.7\textwidth]{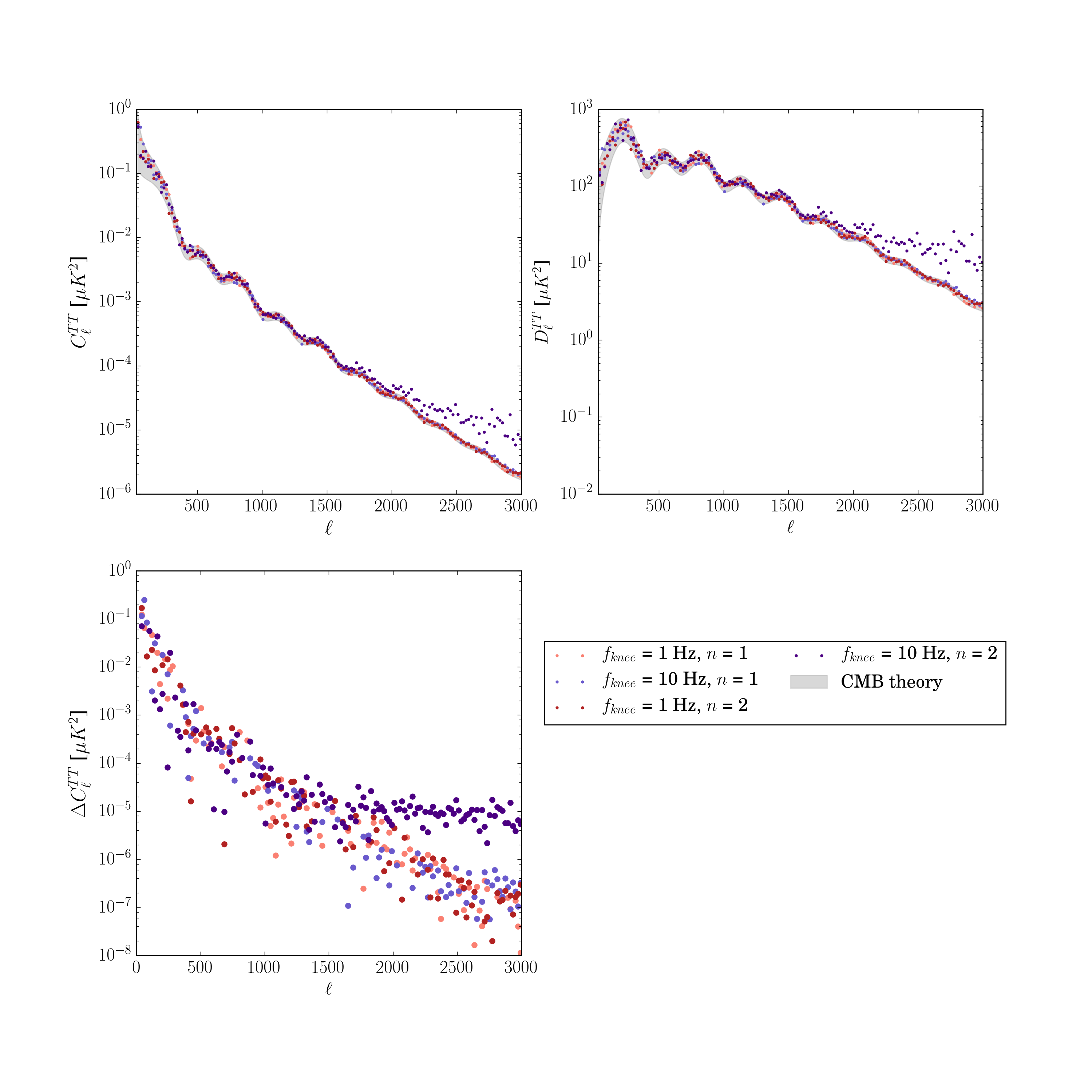}
\caption{Power spectra of CMB signal + noise maps. The {\it top left} plot is the temperature in units of true temperature variance of $\mu K^2$. The {\it bottom left} plot is derived from the one above with the theoretical pure CMB power being subtracted, i.e., it is a measure of the noise bias. The {\it top right} plot is similar to the {\it top left} plot, the scale $D_{\ell} = C_{\ell} \frac{\ell(\ell+1)}{2\pi}$ is a canonical representation of CMB spectra. In these plots the gray-band is cosmic variance, i.e., the statistical uncertainty due to the small sky-fractions simulated here. }
\end{center}
\label{Fig:Cells}
\end{figure}

Ultimately we wish to analyze the CMB in presence of such noise. Thus simply adding, pixel-by-pixel, the CMB signal to such noise maps generate the signal+noise map. This allows us to see how the CMB power spectrum is affected by the variations of the noise PSD, and this is shown in Fig.~\ref{Fig:Cells} for the four noise PSDs considered here, namely with $n = 1, \; 2$ and $f_{knee} = 1, \;10$ Hz. We note here that all noise models introduce varying levels of spectral tilt. The most acute case is for $f_{knee}$ is 10 Hz and high 1/f noise power $n = 2$, where enough noise enters the angles scanned such that it appears as elevated white-noise level, this is an extreme case of a poor detector.

\section{Results and conclusion}
The CMB maps are parametrized as function of cosmological parameters (dark matter density, hubble constant etc.), and assuming some crude noise level, the data maps are analyzed to infer the best fit cosmological parameters. Here we have introduced detector level parameters $n,\, f_{knee}$ and in this process we see how detector noise spectra might affect the CMB spectra as shown in Fig.~\ref{Fig:Cells}. Thus analyzing these maps provides a mapping between parameters obtained from laboratory characterization of detectors and the cosmological parameters that will be obtained from designing a telescope with such detectors; this is the main result of this work.\\

\begin{figure}[h!!]
\begin{center}
\includegraphics[width=0.71\textwidth]{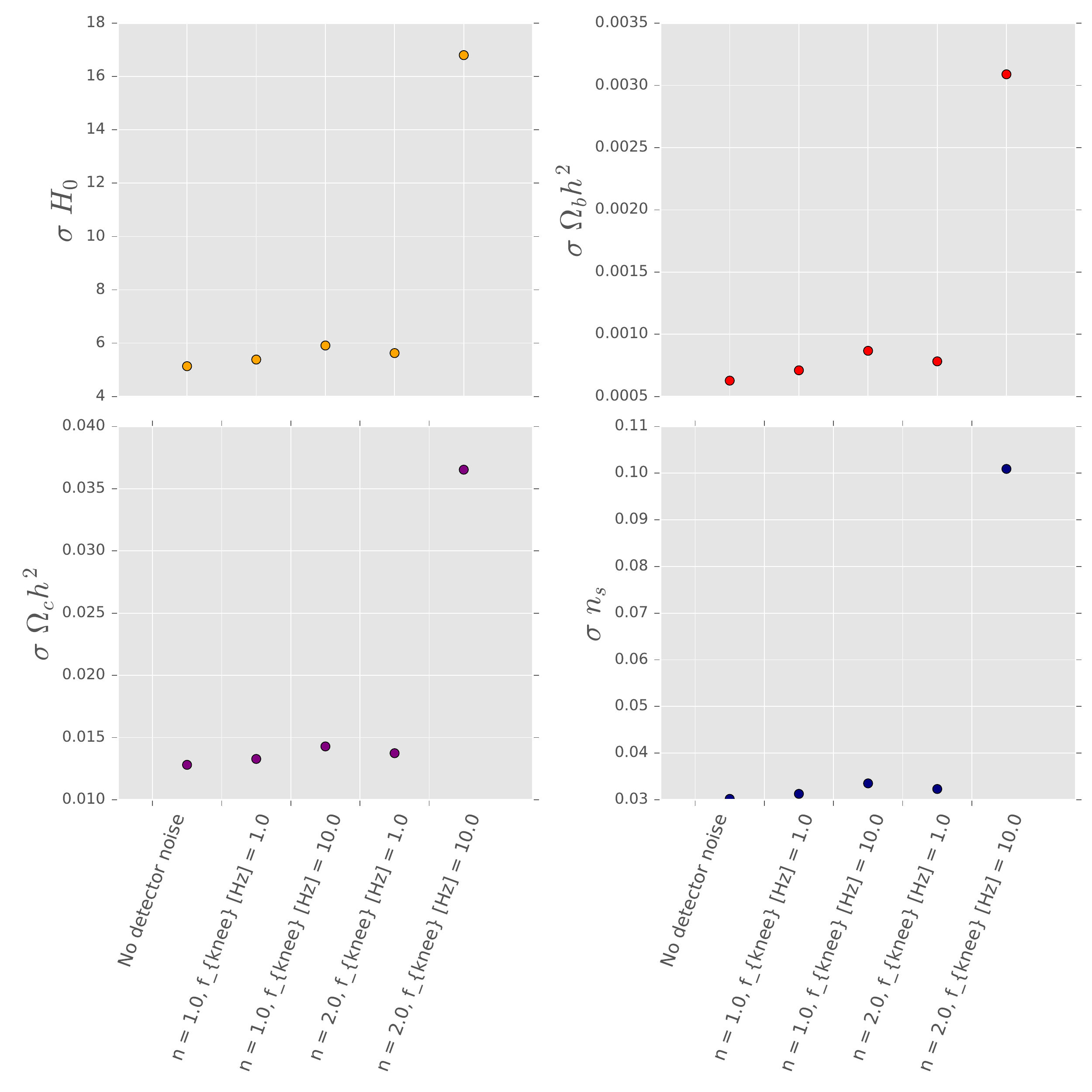}
\caption{Errors in cosmological parameters, Hubble constant $H_0$, baryon and dark-matter densities $\Omega_bh^2$ and $\Omega_ch^2$ and scalar spectral index $n_s$, as function of the two noise parameters $n,\, f_{knee}$. using Fisher matrices.  }
\end{center}
\label{Fig:finale}
\end{figure}

For an explicit parameter mapping example we consider the error on four cosmological parameters, Hubble constant $H_0$, the density of baryons and dark-matter $\Omega_bh^2$ and $\Omega_ch^2$ respectively and the scalar spectral index $n_s$, as function of the two noise parameters $n,\, f_{knee}$. The results are shown in Fig.~\ref{Fig:finale}. From Fig.~\ref{Fig:Cells} we indeed expect the acutely poor PSD ($n, f_{knee} =$ 2, 10 Hz) to have large noise biases. An interesting observation from this simple analysis is that the {\it knee} of the low-frequency noise is more influential than the slope of the noise in affecting the error of the inferred cosmological parameters. From Fig.~\ref{Fig:Cells} we also expect the best-fit values to be systematically affected. The central value estimation is computationally intensive and is beyond the scope of this proceeding.

From Fig.~\ref{Fig:finale} we note that for CMB-S4 like precision surveys the noise power spectrum is important to understand while designing telescopes. We have constructed a framework that allows one to directly take laboratory measurements of detector noise and forward model the effects on the science cases. This is easily done by replacing our model PSDs with measured PSDs for various detectors under consideration. In this proceeding we focused on the low-frequency characteristics of KID noise to show this framework in action. In principle other experimental aspects such as detector responsivity, beam-profile, atmospheric-noise etc. can be easily incorporated to obtain further levels of sophistication for future experiments. 

\begin{acknowledgements}
RBT and JH are supported at UChicago by the Kavli Institute for Cosmological Physics through grant NSF PHY-1125897 and an endowment from the Kavli Foundation and its founder Fred Kavli. JH is supported by the NSF under Award No. AST-1402161. This work was also supported in part by the US DOE under contract No. DE-AC02-76SF00515. We are very thankful to Brad Benson and Clarence Chang for feedback on this work.
\end{acknowledgements}


\begin{thebibliography}{99}

\bibitem{S4}
CMB-S4 Technology Book, First Edition, CMB-S4 Collaboration (Maximilian H. Abitbol (Columbia U.) et al.), 2017 FERMILAB-FN-1034-AE e-Print: arXiv:1706.02464

\bibitem{SS}
Hailey-Dunsheath, S., Shirokoff, E., Barry, P.S. et al. {\it J. Low Temp. Phys.} (2016) \textbf{184}: 180. https://doi.org/10.1007/s10909-015-1375-x

\bibitem{SS2}
Hailey-Dunsheath, S., Shirokoff, E., Barry, P. S., Bradford, C. M., Chattopadhyay, G., Day, P., Zmuidzinas, J. (2014). Status of SuperSpec: A broadband, on-chip millimeter-wave spectrometer. In Proceedings of SPIE - The International Society for Optical Engineering (Vol. 9153). [91530M] SPIE. DOI: 10.1117/12.2057229

\bibitem{NK2}
A. Monfardini et al 2011 ApJS \textbf{194} 24 https://doi.org/10.1088/0067-0049/194/2/24

\bibitem{BLAST}
S.~Gordon {\it et al.},
  J.\ Astron.\ Inst.\  {\bf 05}, no. 04, 1641003 (2017)
  doi:10.1142/S2251171716410038
  arXiv:1611.05400

\bibitem{Yates} 
  S.~J.~C.~Yates, J.~J.~A.~Baselmans, A.~Endo, R.~M.~J.~Janssen, L.~Ferrari, P.~Diener and A.~M.~Baryshev,
  Appl.\ Phys.\ Lett.\  {\bf 99}, 073505 (2011)
  doi:10.1063/1.3624846
  arXiv:1107.4330 

\end{thebibliography}
\end{document}